\documentclass[prd,aps,floats,twocolumn]{revtex4}
\usepackage[usenames, dvipsnames]{color}
 \usepackage{graphicx, array}
\usepackage{graphics,epsfig}
 \usepackage{amssymb}
 \usepackage{amsmath}
 \usepackage{ulem}
 \usepackage{multirow}
  \usepackage{subfigure}
  \usepackage{ulem}

 \def\be{\begin{equation}}
\def\ee{\end{equation}}
 \def\bi{\begin{itemize}}
 \def\ei{\end{itemize}}
  \def\ben{\begin{enumerate}}
\def\een{\end{enumerate}}
  \def\bt{\begin{tabular}}
\def\et{\end{tabular}}
\def\bc{\begin{center}}
\def\ec{\end{center}}

\def\bea{\begin{eqnarray}}
\def\eea{\end{eqnarray}}

\def\ba{\begin{eqnarray}}
\def\ea{\end{eqnarray}}

\def\Vij{V}

\begin{document}

\input{epsf}

\title{Constraints on massive neutrinos from the pairwise kinematic Sunyaev-Zel'dovich effect}
\author {Eva-Maria Mueller$^{1,2}$, Francesco de Bernardis$^{1}$, Rachel Bean$^{2}$, Michael D. Niemack$^{1}$}
\affiliation{ $^1$ Department of Physics, Cornell University, Ithaca, NY 14853, USA,
\\
$^2$ Department of Astronomy, Cornell University, Ithaca, NY 14853, USA.}
\label{firstpage}

\begin{abstract}
We present the mean pairwise momentum of clusters, as observed through the kinematic Sunvaev-Zel'dovich (kSZ) effect, as a novel probe of massive neutrinos. We find that kSZ measurements with current and upcoming surveys will provide complementary constraints on the sum of neutrino masses from large scale structure and will improve on Planck satellite measurements of the primordial cosmic microwave background (CMB) and CMB lensing. 
Central to the constraints is a distinctive scale dependency of the kSZ neutrino signature on the mean pairwise momentum of clusters that we do not expect to be mirrored in systematic effects that change the overall amplitude of the signal, like the cluster optical depth. 
 Assuming a minimal $\Lambda$CDM cosmology including massive neutrinos with Planck primordial CMB priors combined with conservative kSZ specifications, we forecast 68\% upper limits on the neutrino mass sum of $290$ meV, $220$ meV, $96$ meV for ``Stage II'' (ACTPol + BOSS), ``Stage III'' (Advanced ACTPol + BOSS), and ``Stage IV'' (CMB-S4 + DESI) surveys respectively, compared to the Planck alone forecast of 540 meV.  
These forecasts include the ability to simultaneously constrain the neutrino mass sum and the mass-averaged optical depth of the cluster sample in each redshift bin. 
If the averaged optical depth of clusters can be measured with few percent accuracy and a lower limiting mass is assumed, the projected kSZ constraints improve further  to $120$ meV, $90$ meV and $33$ meV (Stage II, III and IV). These forecasts represent a conservative estimate of neutrino constraints using cross-correlations of arcminute-resolution CMB measurements and spectroscopic galaxy surveys. More information relevant for neutrino constraints is available from these surveys, such as galaxy clustering, weak lensing, and CMB temperature and polarization.

\end{abstract}

\maketitle

\section{Introduction}
\label{sec:intro}
Neutrino flavor oscillations in atmospheric and solar neutrinos suggest that neutrinos are massive, implying an extension to the original Standard model of particle physics (e.g., \cite{Fukuda:1998mi, Ashie:2004mr,Ahmad:2001an, Ahmad:2002jz}). These results have also been confirmed by reactor experiments and particle accelerators (e.g., \cite{An:2012eh, Eguchi:2002dm,Ahn:2006zza, Abe:2011sj}). Flavor oscillation experiments, however, cannot measure the neutrino mass eigenstates individually but only the squared mass differences \cite{1694801},
\bea
\Delta m^2_{21}=(7.59\pm0.20)\times 10^{-5} \ \mathrm{eV}^2 \\
|\Delta m_{31}|^2=(2.43\pm0.13)\times 10^{-3} \ \mathrm{eV}^2.
\eea
These measurements allow two distinct neutrino mass hierarchies: normal ($m_1 < m_2 \ll m_3$) and inverted ($m_3 \ll m_1 \simeq m_2$), as well as a degenerate case ($m_3 \simeq m_2 \simeq m_1$). The difference in the squared masses implies there is a minimum neutrino mass sum of $\sum m_\nu \approx 58$ meV for the normal hierarchy, roughly two times larger for the inverted hierarchy, and $\sim150$ meV for the degenerate case \cite{CMBS4neu:2013}.
Cosmological probes are primarily sensitive to the total sum of the neutrino masses, $\sum m_\nu$.
Direct detection experiments in combination with cosmological probes will provide valuable insights into the neutrino hierarchy.

Massive neutrinos affect the background expansion of the universe as well as the growth of structure, which enables constraints from a variety of observations. At early times massive neutrinos are relativistic, acting as a radiation component, though they become non-relativistic at a redshift that scales with the mass. Cosmic microwave background (CMB) experiments can probe massive neutrinos through their effect on the redshift of matter-radiation equality, $z_{\mathrm{eq}}$, or the distance to the last scattering surface, $d_A(z_{\mathrm{dec}})$. Furthermore, since massive neutrinos suppress the matter power spectrum any probe of large scale structure (LSS), such as galaxies and galaxy clusters as well as Lyman-$\alpha$ forest, is potentially sensitive to $\sum m_\nu$.

Current constraints on $\sum m_\nu$ depend strongly on the data sets used. Assuming a 6-parameter minimal flat $\Lambda$CDM cosmology  plus the neutrino mass sum as a free parameter leads to an upper limit of $\sum m_\nu<0.663$ eV at 95\% confidence level for CMB data only, Planck+WMAP+highL (ACT+SPT) \cite{Ade:2013zuv}. Including large scale structure information can further improve the results, yielding $\sum m_\nu<0.230$ eV (95\%, for Planck+WMAP+highL+BAO) \cite{Ade:2013zuv}. Combining more datasets, such as Lyman-$\alpha$ surveys, can lead to even stronger constraints \cite{Palanque-Delabrouille:2014jca}.

Neutrino science is not only interesting from a particle physics perspective, for it has also been used as a means of reconciling observed tensions in the primary CMB anisotropies, cluster counts, and lensing as measured by the Planck satalite \cite{Wyman:2013lza}, and between Planck and the Background Imaging of Cosmic Extragalactic Polarization experiment (BICEP) regarding the tensor-to-scalar ratio \cite{Dvorkin:2014lea}. However, understanding systematic effects, such as cluster mass calibration for the former and foreground contamination by dust emission for the latter, is crucial for interpreting these discrepancies before ascribing them to neutrino physics.  
Nevertheless this illustrates that, despite having relatively small masses, neutrinos can play an important role in cosmology.

In order to reduce systematic effects, measurements from a variety of different cosmological observations are necessary. In this paper we  consider CMB secondary anisotropies, specifically the mean pairwise velocity of clusters as observed through the kinematic Sunyaev Zel'dovich (kSZ) effect \cite{Sunyaev:1980nv}, as a novel probe of neutrino properties. The kSZ effect results in a distortion of the CMB blackbody spectrum caused by CMB photons, passing through clusters, being Doppler shifted due to the line of sight component of the peculiar velocity of the clusters.
Despite its small amplitude, the kSZ effect has been detected from a large sample of clusters by cross-correlating CMB maps with luminous red galaxy (LRG) positions and redshifts \cite{Hand:2012ui}. Previously we studied the potential of using kSZ measurements to test dark energy and modified gravity models \cite{Mueller:2014nsa}. In this work we extend the standard $\Lambda$CDM parameter space to massive neutrinos, parametrized in terms $\sum m_\nu$, and forecast constrains on $\sum m_{\nu}$ from upcoming surveys.

The paper is organized as follows. In Section \ref{sec:Formalism} we: a) give a summary of the effect of massive neutrinos on the power spectrum, halo mass function (HMF), and the growth rate, b) present the formalism of the pairwise statistics of clusters in $\Lambda$CDM cosmological models including massive neutrinos, and c) describe modeling of the covariance matrix and nuisance parameters as well as survey assumptions. In Section \ref{sec:analysis} we describe our analysis and present results before concluding in Section \ref{sec:conclusions}.

\section{Formalism}
\label{sec:Formalism}

\subsection{Cosmic structure and massive neutrinos}
\label{sec:MN_LSS}
We provide a short summary of the cosmological effects of massive neutrinos, focusing on the effect of massive neutrinos on the LSS, and point the reader to \cite{Lesgourgues:2006nd, Lesgourgues:2014zoa, 2011APh....35..177A} as helpful reviews for more details.

Neutrinos cannot be confined on scales below their free-streaming scale, $k_{\mathrm{fs}}$, given by
\bea
k_{\mathrm{fs}}(z) = 0.8 \frac{\Omega_\Lambda + \Omega_m(1+z)^3}{(1+z)^2} \left(\frac{m_\nu}{1\mathrm{eV}}\right) h \  \mathrm{Mpc}^{-1}
\eea
where $\Omega_m$ and $\Omega_\Lambda$ are the matter and $\Lambda$ energy density of today and therefore suppress the matter power spectrum for $k>k_{\mathrm{fs}}$.
Note that different neutrino masses have different free-streaming wavenumbers, leaving LSS not only sensitive to the total neutrino mass but potentially also to individual masses. We calculate the matter power spectrum using CAMB \cite{Lewis:1999bs} taking the effect of massive neutrinos into account.

The universality of the halo mass function in the context of massive neutrino cosmologies has been studied in detail (e.g., \cite{Brandbyge:2010ge,Castorina:2013wga}).
It was found that the halo mass function as well as the halo bias is more accurately described if only the cold dark matter and baryonic matter components, $\Omega_m-\Omega_\nu$, are taken into account when calculating the halo mass function rather than the total mass, $\Omega_m=\Omega_{\mathrm{cdm}}+\Omega_b+\Omega_\nu$. Following \cite{Castorina:2013wga} we will denote the cold dark matter and baryonic component as ``cold" matter (subscript ``c") in contrast to total matter that includes the neutrino mass (subscript ``m"). The comoving number density of halos per unit mass, $dn(M)/dM$, can then be modeled in the common form
\bea
\frac{dn(M,z)}{dM}=\frac{\rho_c}{M} f(\sigma_c,z)\frac{d\mathrm{ln}\sigma_c^{-1}}{dM},
\eea
using $\rho_c=\rho_{\mathrm{cdm}}+\rho_b$ as the background matter density and with the r.m.s of the linear cold matter density field given by
\bea
\sigma_c^2(M,z)=\int d^3kP^{cc}_{\mathrm{lin}}(k,z)W^2(Rk),
\eea
where $P^{cc}_{\mathrm{lin}}$ is the linear cold matter power spectrum and the scale R is given by $M=\frac{4\pi}{3}\rho_c R^3$.
Many different fitting functions have been proposed within the literature, however, for the purpose of this work the choice of halo mass function has little effect. Here we adopt the fitting function given in \cite{2011ApJ...732..122B} to parametrize the halo multiplicity function, $f(\sigma_c,z)$.

Similarly we define the halo bias with respect to the cold matter, $b_c(M,z)$, using the description given in \cite{2011ApJ...732..122B}.

The growth rate, $f_g$, is defined as
\bea
f_g(a) \equiv \frac{d\mathrm{ln}D(a)}{d\mathrm{ln}a},
\eea
with the growth factor, $D(a)$, normalized to $D(a_0)=1$, and becomes scale dependent in the presence of massive neutrinos.
Here we use the fitting function
\bea
f_g(z,k) \approx \mu(k)\Omega_m^{\gamma}(z), \label{eq:fg}
\eea
where
\bea
\mu(k) = 1-A(k)\Omega_\Lambda f_\nu + B(k) f_\nu^2-C(k)f_\nu^3 \label{eq:mu}
\eea
with $A(k)$, $B(k)$, $C(k)$ given in Table II of \cite{Kiakotou:2007pz} and with the growth exponent, $\gamma$, parametrized as $\gamma=0.55+0.05[1+w(z=1)]$ \cite{Linder:2005in}.

\subsection{Motion of clusters as a probe of massive neutrinos}

\label{sec:Vel}
The mean pairswise velocity of clusters can be modeled \cite{Sheth:2000ff, Bhattacharya:2007sk, Mueller:2014nsa} as
\bea
\Vij(r,a)&=&-\frac{2}{3} H(a) a \ \Omega_m^\gamma(a)\frac{r \bar{\xi}_{h}(r,a)}{1+\xi_{h}(r,a)}, \label{eq:massavvij}
\eea
with $\xi_h$ and $\bar{\xi}_h$ given by
\bea
\xi_{h}(r,a)&=&\frac{1}{2 \pi^2}\int dk k^2 j_0(kr) P_{\mathrm{lin}}(k,a) b_{h,c}^{(2)}(k), \hspace{0.5cm}
\\
\bar{\xi}_{h}(r,a)&=&\frac{3}{ r^3} \int_0^{r} dr' r'^2 \mu(k,a) \xi(r',a) b_{h,c}^{(1)}(k),
\eea
where $P_{\mathrm{lin}}$ is the linear, total matter power spectrum and $b_{h,c}^{(q)}$ are the moments of the halo bias defined as
\bea
b^{(q)}_{h,c}&=&\frac{\int dM \ M \ n(M,z)b^{q}_c(M,z)W^2[kR]}{\int dM\ M \ n(M,z)W^2[kR]}. \hspace{0.5cm} \label{eq:bq}
\eea
Note that the moments of the halo bias are taken with respect to the halo mass function and halo bias, including the effect of massive neutrinos as described above.

\begin{figure}[!t]
\bc
{\includegraphics[width=0.49\textwidth]{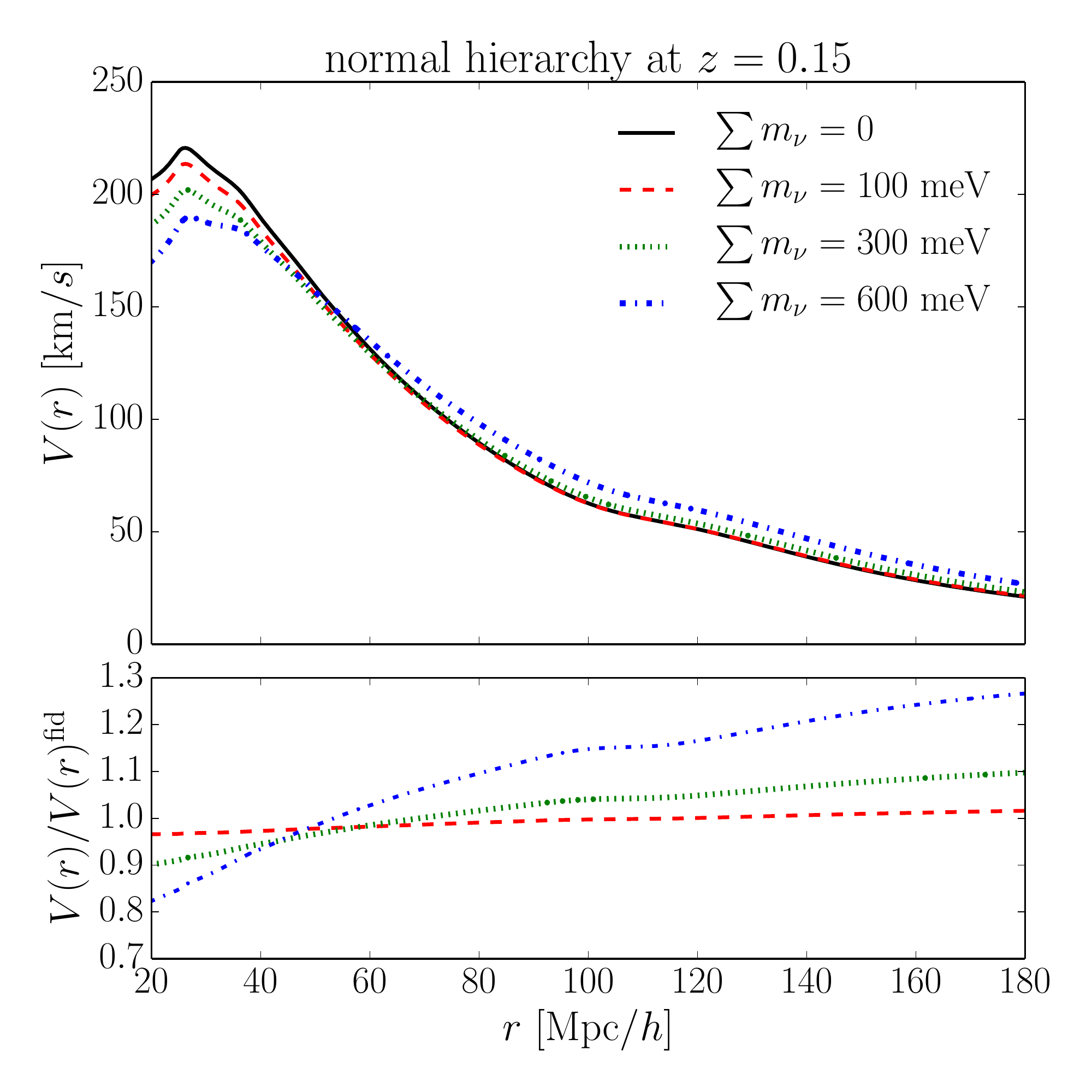}
}
\caption{[Upper panel] The mean pairwise cluster velocity, $\Vij(r)$, for different neutrino mass sums, $\sum m_\nu$. $\Vij(r)$ is calculated at $z=0.15$ and assuming a minimum cluster mass of $M_{\mathrm{min}}=1\times 10^{14} M_\odot$ with other cosmological parameters, including $\Omega_m h^2$, fixed to fiducial values. Here we assume a normal neutrino hierarchy ($m_1\approx m_2\approx 0, m_3\neq 0$). [Lower panel] Ratio of the mean pairwise velocity, $\Vij (r)$, for the different scenarios to the fiducial model with $\sum m_\nu = 0$ (black line in upper panel). Massive neutrinos clearly leave a scale dependent imprint on $\Vij(r)$, which is highlighted by the different slopes of the lower curves.}
\label{fig:Vij_mass}
\ec
\end{figure}

\begin{figure}[!t]
\bc
{\includegraphics[width=0.49\textwidth]{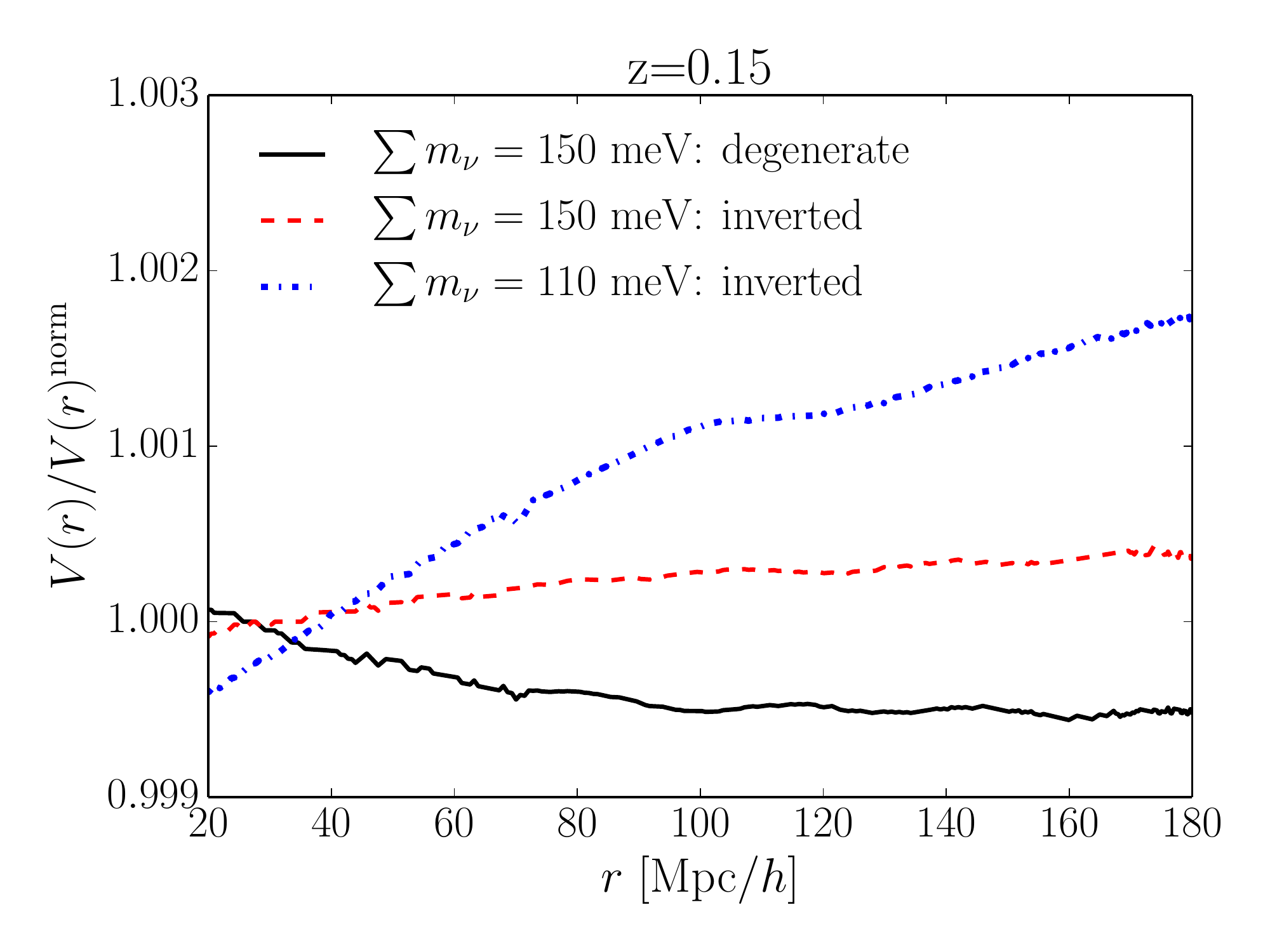}
}
\caption{Ratio of the mean pairwise velocity, $\Vij(r)$, for the different neutrino models to the normal hierarchy ($m_1 < m_2 \ll m_3$) , $\Vij(r)^{norm}$.  The inverted hierachy ($m_3 \ll m_1 \simeq m_2$) deviates more strongly from the normal hierarchy assuming a total mass of $\sum m_\nu=110$ meV (blue, dashed-dotted line) than for $\sum m_\nu=150$ meV (red, dashed line). A degenerate model ($m_1\approx m_2 \approx m_3$, black, solid line) with $\sum m_\nu=150$ meV exhibits a slightly different behavior from the normal and inverted hierarchies. This shows that there is a mild sensitivity to the different masses of the individual neutrinos even when $\sum m_\nu$ is constant. }
\label{fig:Vij_hier}
\ec
\end{figure}

Figure \ref{fig:Vij_mass} shows the mean pairwise velocity of clusters for different neutrino masses assuming a normal hierarchy. The mass of the neutrinos leave an $r$-dependent imprint on $V(r)$, which enables us to differentiate between the mass of the neutrinos and other cosmological parameters and systematic effects that primarily change the overall amplitude of $V(r)$. Figure \ref{fig:Vij_hier} displays the mild sensitivity of $V(r)$ to the different neutrino hierarchies.

\subsection{General methodology and survey assumptions}
\label{sec:Cov}
We  use the Fisher matrix approach to estimate the covariance matrix between parameters $p_\mu$ and $p_\nu$, $C_{\mu\nu}=F_{\mu\nu}^{-1}$, with
\bea
F_{\mu \nu}=\sum_{i}^{N_z}\sum_{p,q}^{N_r} \frac{\partial V(r_p,z_i)}{\partial p_{\mu}} \mathrm{Cov}^{-1}_i(r_p,r_q,z_i) \frac{\partial V(r_p,z_i)}{\partial p_{\nu}},
\eea
where $\mathrm{Cov}(r_p,r_q,z_i)$ is the covariance matrix, as defined in \cite{Mueller:2014nsa}, between two clusters pairs at redshift $z_i$ with each pair having comoving separations of $r_p$ and $r_q$,  including cosmic variance, shot noise and a measurement uncertainty due to the sensitivity of the CMB instrument (given in \cite{Mueller:2014nsa} Table III). $N_z$ and $N_r$ are the number of redshift and spatial separation bins, respectively. 

The kSZ effect is proportional to the mean pairwise momentum of clusters, $\sim\tau v$.  In order to accurately measure cluster velocities, the average cluster optical depth and its dependency on mass and redshift must be known. The optical depth can be modeled, for instance using hydrodynamical simulations \cite{Battaglia:2010tm}, or constrained using complementary cluster data such as CMB polarization  \cite{Sazonov:1999zp}.  
We allow large uncertainties in the cluster optical depth that could introduce systematic biases in $\Vij(r)$ by marginalizing over a potential bias, defined via $\hat{V}(z_i) = b_\tau(z_i) V(z_i)$, that scales the amplitude in each redshift slice. 
We also introduce the limiting cluster mass of the sample as a nuisance parameter in our analysis, imposing a 15\% prior.

We  consider two sets of cosmological parameters. The first is a minimal 6-parameter flat $\Lambda$CDM cosmological model plus the neutrino mass sum and (1+$N_z$) nuisance parameters, which we denote as $\Lambda$M(ixed)DM and summarize as
\bea
\bold{p}_{\mathrm{min}}&=&\{\Omega_b h^2,\Omega_m h^2,\Omega_\Lambda,n_s, \mathrm{log}(10^{10}A_s),\sum m_\nu\}  \nonumber \\
&& + \{M_{\mathrm{min}}, b_\tau(z)\},
\eea
where $\Omega_b$, $\Omega_m$, $\Omega_\Lambda$ are the dimensionless baryon, matter and dark energy densities respectively, $h$ is the Hubble constant in units of 100 km/s/Mpc, $n_s$ and $A_s$ are the spectral index and normalization of the primordial spectrum of curvature perturbations, $M_{\mathrm{min}}$ is the limiting cluster mass of the catalog, and $b_\tau$ the nuisance parameter due to uncertainty in $\tau$.  The matter energy density includes cold dark matter, baryons, and the neutrino mass contribution, i.e. $\Omega_m=\Omega_{\mathrm{cdm}}+\Omega_b+\Omega_\nu$, where $\Omega_\nu$ is the neutrino energy density, related to the neutrino mass via $\Omega_\nu = \sum m_\nu/(93.14\ h^2\mathrm{eV})$ (see e.g., \cite{Lesgourgues:2014zoa}).  
The second set of parameters is a more general $\Lambda$MDM model that also includes the curvature energy density, $\Omega_k$, and the dark energy equation of state parameters, $w_0$ and $w_a$ such that the equation of state is $w(a)=w_0+(1-a)w_a$, or 
\bea
\bold{p}=\bold{p}_{\mathrm{min}}+\{\Omega_k, w_0,w_a\}.
\eea

We adopt the fiducial values assumed in the Euclid Assessment Study report \cite{Laureijs:2011gra},
$ \omega_b=0.021805, \omega_m=0.1225, \Omega_\Lambda=0.75, n_s=1, \mathrm{log}(10^{10}A_s)= 3.2336, \Omega_k=0, w_0=-0.95,w_a=0 $ which corresponds to $\Omega_m=0.25,\Omega_b=0.0445,h=0.7$ and $\sigma_8=0.8$. For the CMB Fisher matrix we also marginalize over the reionization optical depth with fiducial value $\tau_0 = 0.11$.
We assume a normal hierarchy ($m_1\approx m_2\approx 0, m_3\neq 0$) as our reference case with the fiducial total neutrino mass, $\sum m_\nu=60$ meV. We find that our results are robust to the assumed fiducial cosmology, including the fiducial neutrino mass, and only show a mild sensitivity to the assumed neutrino hierarchy. 

The kSZ effect can be extracted by cross-correlating the CMB surveys with cluster positions and redshifts, using LRGs as a tracer for clusters or by using a photometrical selected cluster catalog such as RedMApper \cite{Rykoff:2013ovv}.
In this work we consider three potential surveys: Stage II -- ACTPol \cite{Niemack:2010wz} cross-correlated with BOSS \cite{Dawson:2012va}, Stage III -- Advanced ACTPol \cite{Calabrese:2014gwa} and BOSS, and Stage IV -- CMB Stage IV \cite{CMBS4neu:2013} and DESI \cite{Levi:2013gra}. Details of the survey assumptions, including redshift range, overlapping sky coverage between CMB and LSS surveys, and the minimum cluster mass of the potential cluster catalogs, are summarized in Table \ref{tab:kSZ_surveys}.  

\begin{table}[!t]
\begin{center}
	\begin{tabular}{|l|l|c|c|c|}
	\cline{3-5}
	 \multicolumn{2}{c}{} & \multicolumn{3}{|c|}{ Survey Stage }
\\	\hline
 Survey & Parameters \  & \ II \ & \ III \ & \  IV \    \\
    \hline
 CMB&    $\Delta T_{\mathrm{instr}}$ ($\mu K \mathrm{arcmin}$) \ & 20 &  7  & 1
    \\ \hline
 \multirow{4}{*}{Galaxy} \ &   $z_{\mathrm{min}}$ &0.1&0.1&0.1
    \\
   &  $z_{\mathrm{max}}$  \ &0.4&0.4&0.6
    \\
& No. of $z$ bins, $N_z$   \   &3&3&5
    \\
  &  $M_{\mathrm{min}}$ ($10^{14}M_\odot $) \ & $1 $ & $1$ &  $0.6$
  \\  \hline
  Overlap &    Area (1000 sq. deg.) \ & 4 & 6& 10
    \\ \hline
    \end{tabular}
    \caption{Reference survey specifications used to model Stage II, III and IV  
    kSZ cluster surveys. The expected instrument sensitivity of the CMB survey, $\Delta T_{\mathrm{inst}}$, along with the assumed optical large scale structure survey redshift range $z_{\mathrm{min}}<z<z_{\mathrm{max}}$,  redshift binning, and minimum detectable cluster mass, $M_{\mathrm{min}}$ are shown. We consider an effective sky coverage 
    by estimating the degree of overlap between the respective CMB and optically selected cluster datasets.}
    \label{tab:kSZ_surveys}
\end{center}
\end{table}

We are conservative in our survey specifications by assuming a photometrical selected cluster catalog, to ensure completeness and purity of our sample,  with spectroscopically selected LRGs to give cluster redshifts. We also only assume single frequency, 150 GHz, CMB measurements (even though e.g., Advanced ACTPol will have five frequency bands). 
The assumed sky coverage for the kSZ analysis is subsequently limited by the overlapping area of photometric and spectroscopic LSS surveys with CMB kSZ measurements. The assumed redshift range is given by that for the spectroscopic LRG sample, and the limiting mass is motivated by the expected, photometrically selected cluster catalogs. A more detailed discussion can be found in \cite{Mueller:2014nsa}.

In addition to the fiducial scenario, we consider a more optimistic lower cluster mass limit of $M_{\mathrm{min}}=4\times 10^{13} M_\odot$ for Stage II, III and $M_{\mathrm{min}}=1\times 10^{13} M_\odot$ for Stage IV.
For our CMB priors we use the survey specification for a Planck-like survey, including primordial temperature and polarization as well as lensing information,  as given in \cite{Lesgourgues:2005yv}. 
\section{Analysis }
\label{sec:analysis}

\subsection{Potential kSZ constraints on massive neutrinos}
\label{sec:forecast_summary}

\begin{figure*}[!thb]
\bc
{\includegraphics[width=\linewidth]{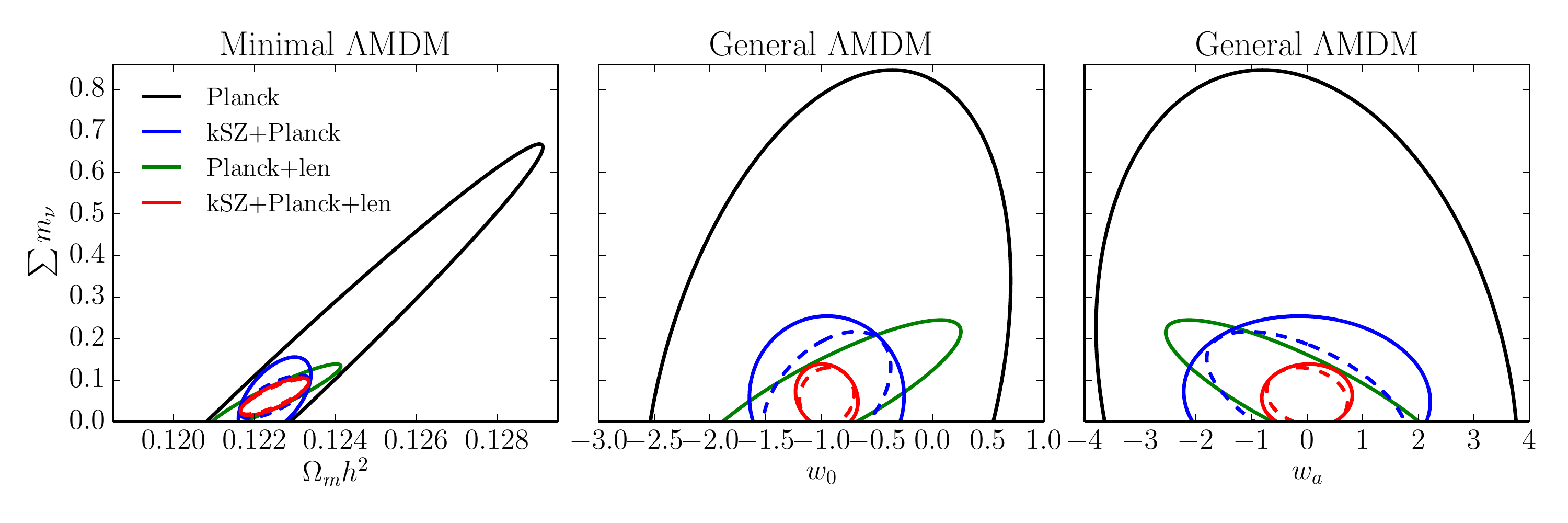}
}
\caption{Constraints on the neutrino mass sum, $\sum m_\nu$, versus  the matter density, $\Omega_m h^2$ (left), and the dark energy equation of state, $w_0$ (middle), and its evolution, $w_a$ (right). Contours show 2D 68\% confidence levels for the following data sets: Planck (black), Planck and kSZ Stage IV (blue), Planck including CMB lensing (green), and Planck including CMB lensing plus kSZ  Stage IV(red). Solid lines assume no prior on the average cluster optical depth evolution, $b_\tau(z)$, while dashed lines assume a $1\%$ prior on $b_\tau(z)$. 
}
\label{fig:kSZPlanck_2D}
\ec
\end{figure*}

\begin{table*}[!htb]
\begin{center}
\begin{tabular}{|l|l|r|r|r|| r|r|r|| r|r|r||  }
\cline{3-11}
\multicolumn{2}{c}{} &\multicolumn{9}{|c||}{$\sigma(\sum m_\nu) $ in  meV} \\
\cline{3-11}
\multicolumn{2}{c}{} & \multicolumn{3}{|c||}{no prior on $b_\tau$}&\multicolumn{3}{|c||}{no prior on $b_\tau$, opt. $M_{\mathrm{min}}$ }&\multicolumn{3}{|c||}{ 1\% prior on $b_\tau$, opt. $M_{\mathrm{min}}$}\\
\cline{3-11}
\multicolumn{2}{c|}{}   & Stage II & Stage III & Stage IV & Stage II &   Stage III &  Stage IV & Stage II &   Stage III &  Stage IV \\
\hline
\multirow{2}{*}{kSZ+ Planck}&minimal $\Lambda$MDM &       290 &        220 &       96 &       180 &        140 &       59 &       120 &        90 &       33   \\
 \cline{2-11}
 &+$\Omega_k,w_0,w_a$&       500 &        430 &       200  &       410 &        320 &       110 &       230 &        210 &       98 \\
  \cline{2-11}
  \hline
      \multirow{2}{*}{kSZ+ Planck+lens}&minimal $\Lambda$MDM &       73 &        69 &       46 &      65 &        58  &       30  &       65 &        57 &      28    \\
 \cline{2-11}
 &+$\Omega_k,w_0,w_a$&       86  &        84 &       79 &       84 &        83 &       67&      82 &        79 &       61 \\
  \cline{2-11}
  \hline

\end{tabular}
\caption{ Forecasts for $1\sigma$ errors on $\sum m_\nu$ for a normal hierarchy. [Left columns] Results marginalizing over $b_\tau$ [top row] marginalized constraints on the sum of the neutrinos mass for the kSZ surveys including Planck primordial CMB priors for a 5 parameter minimal $\Lambda$MDM cosmology $+\sum m_\nu$ as well as [second row] including $\Omega_k,w_0,w_a$; and [lower rows] similar results also including Planck lensing constraints.  [Middle columns] Assuming a more optimistic optically-selected mass cutoff of $M_{\mathrm{min}}=4\times 10^{13} M_\odot$ for Stage II and III and  $M_{\mathrm{min}}=1\times 10^{13} M_\odot$ for Stage IV, relative to the fiducial scenarios of $M_{\mathrm{min}}=10^{14}M_\odot$ (Stages II and III) and $M_{\mathrm{min}}=6\times 10^{13}M_\odot$ (Stage IV).  [Right columns] Including a $1\%$ prior on $b_\tau$ and optimistic assumptions on the minimum mass. The CMB Planck priors alone provide $\sigma(\sum m_\nu)=540 \ (660)$ meV for a minimal (general) $\Lambda$MDM cosmology and  $\sigma(\sum m_\nu)=200 \ (240$) meV including CMB lensing.}
\label{tab:results}
\end{center}
\end{table*}

The results of our Fisher matrix analysis, the forecast 1-$\sigma$ errors on the total neutrino mass $\sum m_\nu$, are displayed in Table \ref{tab:results} for our fiducial case as well as more optimistic assumptions on the $\tau$ bias parameter and limiting mass. The left hand columns of Table \ref{tab:results} depict the most conservative scenario in which we treat the mass-averaged optical depth of clusters $\tau$ as an unknown nuisance parameter in each redshift bin, $b_\tau(z)$, that scales the overall amplitude of the mean pairwise velocity, $V(r)$. 
Under this assumption the optical depth of clusters can scale the overall amplitude of $V(r)$ but has no $r$-dependent effect. The scenarios depicted in the middle and right columns show less conservative assumptions on the limiting mass (middle) as well as adding a $1\%$ prior on $b_\tau(z)$ (right). 

With kSZ measurements alone, the total neutrino mass is degenerate with other cosmological parameters, in particular with the matter density $\Omega_m h^2$ and the dark energy equation of state parameters $w_0$ and $w_a$. 
 Since kSZ data alone cannot constrain the cosmological background parameters well \cite{Mueller:2014nsa}, the predicted error on the total neutrino mass are comparatively loose. The forecasted 68\% upper limit on the total neutrino mass with kSZ data only is 1300 meV, 810 meV and 320 meV for Stage II, III and IV respectively for the minimal $\Lambda$MDM cosmology in our most conservative scenario. In comparison, our projected Planck constraints are 540 meV using the primordial temperature and polarization power spectrum and 200 meV when including lensing.

Adding primordial CMB data from Planck can improve the constraints significantly by breaking degeneracies with the other parameters. For a minimal $\Lambda$MDM cosmology in combination with Planck CMB results, kSZ measurements can improve constraints on $\sum m_\nu$ and achieve 290 meV, 220 meV and 96 meV at 68\% confidence level for Stage II, III and IV with no prior on $b_\tau(z)$. Marginalizing over the overall amplitude in each redshift bin does not deteriorate the constraints significantly. This can be understood considering the scale dependent effect of the neutrino mass on $V(r)$, see Figure (\ref{fig:Vij_mass}). The $r$-dependent change of $V(r)$ due to massive neutrinos can not be compensated by an overall shift of the amplitude of $V(r)$; therefore there is not a strong degeneracy between $\sum m_\nu$ and $b_\tau$. Remarkably, even without prior knowledge of the average cluster optical depth, $\tau$, the kSZ data set should significantly improve upon the Planck constraints (see Figure \ref{fig:kSZPlanck_2D}).

The CMB contains additional information related to  $\sum m_\nu$ due to gravitational lensing by large scale structure. We compare the relative potential constraints by including the information from Planck CMB lensing  as a separate case. In Table \ref{tab:results}, {\it Planck} refers to CMB temperature and polarization data only, while {\it Planck+lens} also includes  the forecasts for a Planck CMB lensing extraction performed with a quadratic estimator based on temperature and polarization data \cite{2003PhRvD..67h3002O}. For the minimal $\Lambda$MDM cosmology, when including lensing, the expected errors reduce to 73 meV, 69 meV, 46 meV. As shown in Figure \ref{fig:kSZPlanck_2D}, this represents  a considerable improvement relative to constraints from Planck alone plus CMB lensing. We stress that in this analysis we are assuming that the lensing extraction can be performed perfectly from CMB maps, which is an optimistic assumption. On the other hand, the kSZ constraints include relatively conservative assumptions (see \S \ref{sec:Cov}). Hence, it is likely that the improvement with respect to CMB data alone achievable with pairwise velocities will be even larger than that presented in this paper.

\begin{figure}[!t]
\bc
{\includegraphics[width=0.49\textwidth]{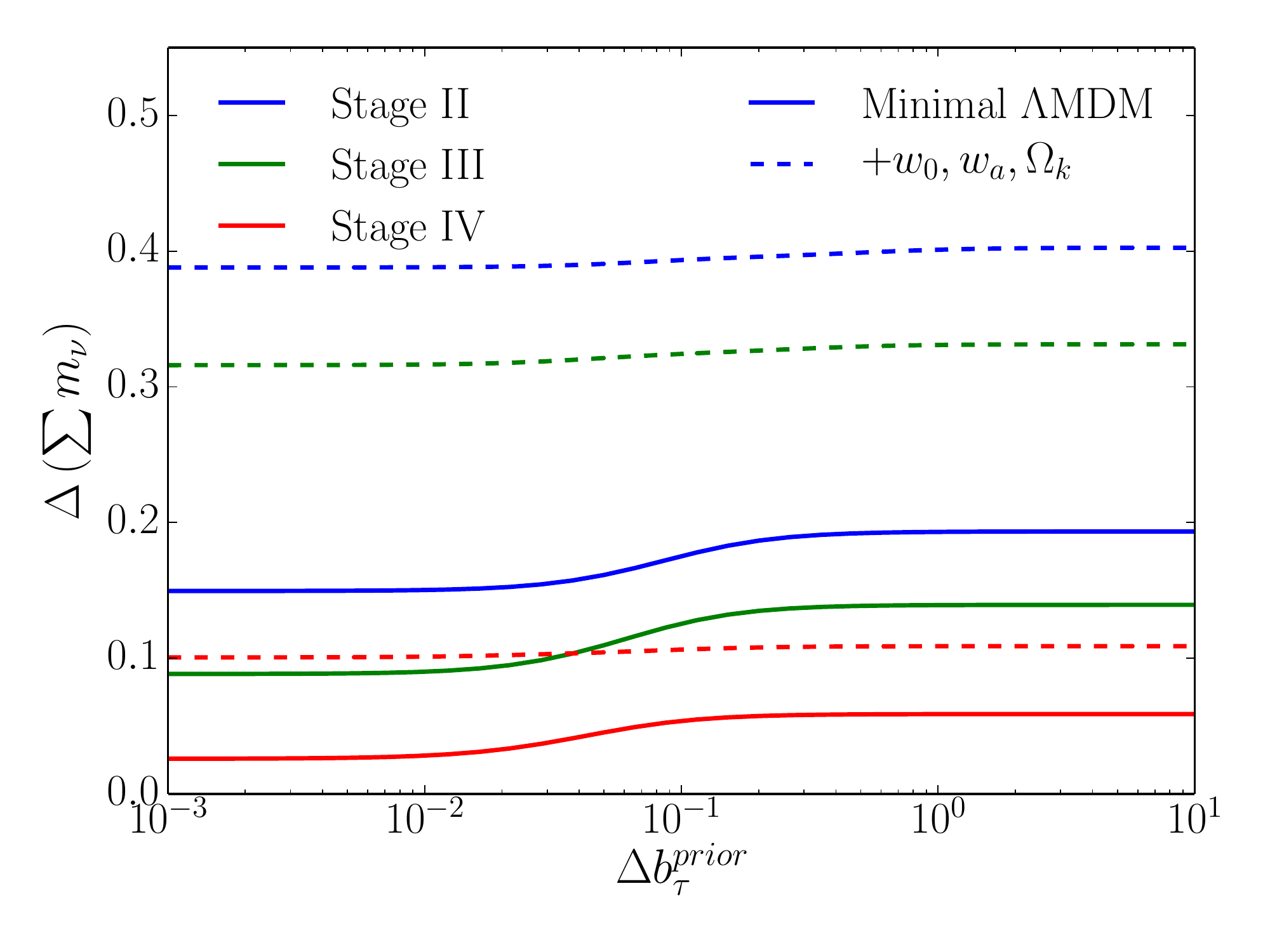}
}
\caption{The impact of imposing a prior on a potential systematic offset in the mass-averaged optical depth, parameterized by a multiplicative bias in each redshift bin, $b_\tau$(z), for Stage II (blue), Stage III (green) and Stage IV (red) surveys, on the 1-$\sigma$ constraints on $\sum m_\nu$. This assumes a mass cutoff of $M_{\mathrm{min}}=4\times 10^{13} M_\odot$ for Stage II and III and  $M_{\mathrm{min}}=1\times 10^{13} M_\odot$ for Stage IV as well as including CMB temperature and polarization priors. This shows that $\sum m_{\nu}$ constraints are more weakly dependent on $\tau$ than the dark energy and gravity parameters studied in \cite{Mueller:2014nsa}. }
\label{fig:dmnu_btau}
\ec
\end{figure}

The right hand columns of Table \ref{tab:results} display the results assuming a more optimistic limiting mass of the cluster sample, $M_{\mathrm{min}}=1\times10^{13}M_\odot$ for Stage IV and $M_{\mathrm{min}}=4\times 10^{13}M_\odot$ for Stage II and III. Under these assumptions, the sum of the total neutrino mass for a Stage IV survey plus CMB priors can be constrained up to 59 meV at $68\%$ confidence level. Imposing a prior on the optical depth bias parameter can further improve the results (see Figure \ref{fig:dmnu_btau}). The uncertainty in $\sum m_\nu$ reduces to 120 meV, 90 meV, 33 eV (Stage II, III and IV) assuming a 1\% prior on the mass-averaged optical depth.

For a more general cosmology including the curvature energy density, $\Omega_k$,  and the dark energy equation of state parameters, $w_0$ and $w_a$, the kSZ measurement constraints are significantly degraded; however, as shown in Table \ref{tab:results}, the constraints remain quite strong when also including Planck lensing measurements to help constrain the background expansion history (see also the discussion in \cite{Mueller:2014nsa}). This highlights the potential power of combining constraints from CMB lensing and kSZ measurements from the same survey instead of relying on the Planck lensing measurements. Here we highlight the sensitivity of kSZ measurements to $\sum m_\nu$ and leave combined lensing and kSZ forecasts to future work.

An important factor that determines the constraining power of kSZ pairwise velocities measurements is the minimum comoving separation achievable by the survey. Photometric surveys would not be able to reconstruct the signal at lower separations as a consequence of the redshift uncertainty. As a test, we have repeated our baseline analysis using a minimum $r$ cutoff $r_{\mathrm{min}}=50$ Mpc/h. This cutoff represents the approximate minimum separation achievable with a photometric survey like the Dark Energy Survey (see, for example, the discussion in \cite{Keisler:2012eg}). In this case, while for kSZ data only the constraints deteriorate more than a factor of two, when including CMB priors from Planck the $1\sigma$ error degradation is only $10-20\%$ depending on the survey considered. For example, for a minimal $\Lambda$MDM model we find a $\sigma(\sum m_{\nu})$ of $76$ meV, $75$ meV and $51$ meV respectively for Stage II, III and IV surveys, from the combination of kSZ data with Planck CMB $+$ lensing with no prior on $b_\tau$(z).

kSZ measurements from combining large scale structure and CMB surveys provide an alternative, complementary approach to constraining the neutrino mass sum from using galaxy cluster and CMB measurements independently. For example, stage IV LRG and emission line galaxy (ELG) clustering forecasts, assuming degenerate massive neutrinos with a total mass $\sum m_\nu=60$ meV, a minimal $\Lambda$MDM cosmology and fiducial values of the cosmological parameters from Planck \cite{Ade:2013zuv}, suggest potential constraints  on the neutrino mass, of 24 meV for DESI \cite{Font-Ribera:2013rwa} including a CMB prior and using the galaxy broadband up to $k_{\mathrm{max}}=0.1$ h/Mpc, after accounting for uncertainties in galaxy bias, shot noise and non-linear damping uncertainties. Similar constraints are forecast for stage IV CMB surveys alone and improved constraints from combining CMB lensing with galaxy surveys (see e.g., \cite{CMBS4neu:2013}). Adding kSZ measurements should improve these constraints for CMB surveys with the roughly arcminute-resolution required for accurate kSZ measurements \cite{Calabrese:2014gwa}.

\section{Conclusions }
\label{sec:conclusions}

The mean pairwise velocity of clusters, as observed through the kSZ effect, is a novel probe of massive neutrinos. Upcoming kSZ measurements are expected to be comparable in their constraining power to upcoming spectroscopic and photometric galaxy surveys and CMB surveys. 
Like other LSS probes, the kSZ cannot differentiate strongly between different hierarchies, but can measure the sum of the neutrino mass through its suppression of the matter power spectrum and its effects in the HMF, halo bias and growth rate, as discussed in Section \ref{sec:Formalism}, leading to a scale dependent variation in $V(r)$. 

Conservative forecasts for kSZ measurements in combination with Planck primordial CMB priors provide constraints on the sum of neutrino masses with a precision of 290 meV, 220 meV and 96 meV for Stage II, III and IV respectively, assuming a fiducial mass of 60 meV, a normal hierarchy and a minimal $\Lambda$MDM cosmology. These constraints include simultaneous constraints on the average optical depth of the cluster sample in different redshift bins by marginalizing over the overall amplitude of $V(r)$ in each bin. The kSZ effect can lead to strong constraints and improve upon Planck results even without precise knowledge of $\tau$, due to the scale dependent effect of massive neutrinos on $V(r)$.
More optimistic assumptions on the limiting mass of the cluster sample and the cluster optical depth can improve the results to 120 meV,  90 meV and 33 meV (Stage II, III and IV). These constraints can be further improved by including CMB lensing measurements.
 
Our conservative analysis has led to potentially powerful constraints on the cosmic growth history to characterize the properties of neutrinos and other cosmological parameters using the kinematic Sunyaev-Zel'dovich effect. Here we have only considered constraints from Planck satellite data combined with kSZ measurements from cross-correlating spectroscopic galaxy surveys and CMB surveys. CMB polarization and X-ray cluster measurements may allow a refined understanding of cluster optical depth. Upcoming measurements of type 1a supernovae, galaxy clustering, weak lensing, and the CMB have the potential to break degeneracies in kSZ measurements, providing complementary constraints on the expansion history, to give even more stringent constraints on the properties of the dark sector including dark energy and neutrinos.

\section*{Acknowledgments}

The authors thank Nicholas Battaglia for helpful discussions on $\tau$ estimation and inputs from simulations that informed the modeling assumptions.
The work of EMM and RB is supported by NASA ATP grants NNX11AI95G and NNX14AH53G, NASA ROSES grant 12-EUCLID12- 0004, NSF CAREER grant 0844825 and DoE grant DE-SC0011838.

\newpage
\bibliographystyle{apsrev}
\bibliography{big_bib_nov19.bib}

\begin{thebibliography}{40}
\expandafter\ifx\csname natexlab\endcsname\relax\def\natexlab#1{#1}\fi
\expandafter\ifx\csname bibnamefont\endcsname\relax
  \def\bibnamefont#1{#1}\fi
\expandafter\ifx\csname bibfnamefont\endcsname\relax
  \def\bibfnamefont#1{#1}\fi
\expandafter\ifx\csname citenamefont\endcsname\relax
  \def\citenamefont#1{#1}\fi
\expandafter\ifx\csname url\endcsname\relax
  \def\url#1{\texttt{#1}}\fi
\expandafter\ifx\csname urlprefix\endcsname\relax\def\urlprefix{URL }\fi
\providecommand{\bibinfo}[2]{#2}
\providecommand{\eprint}[2][]{\url{#2}}

\bibitem[{\citenamefont{Fukuda et~al.}(1998)}]{Fukuda:1998mi}
\bibinfo{author}{\bibfnamefont{Y.}~\bibnamefont{Fukuda}} \bibnamefont{et~al.}
  (\bibinfo{collaboration}{Super-Kamiokande Collaboration}),
  \bibinfo{journal}{Phys.Rev.Lett.} \textbf{\bibinfo{volume}{81}},
  \bibinfo{pages}{1562} (\bibinfo{year}{1998}), \eprint{hep-ex/9807003}.

\bibitem[{\citenamefont{Ashie et~al.}(2004)}]{Ashie:2004mr}
\bibinfo{author}{\bibfnamefont{Y.}~\bibnamefont{Ashie}} \bibnamefont{et~al.}
  (\bibinfo{collaboration}{Super-Kamiokande Collaboration}),
  \bibinfo{journal}{Phys.Rev.Lett.} \textbf{\bibinfo{volume}{93}},
  \bibinfo{pages}{101801} (\bibinfo{year}{2004}), \eprint{hep-ex/0404034}.

\bibitem[{\citenamefont{Ahmad et~al.}(2001)}]{Ahmad:2001an}
\bibinfo{author}{\bibfnamefont{Q.}~\bibnamefont{Ahmad}} \bibnamefont{et~al.}
  (\bibinfo{collaboration}{SNO Collaboration}),
  \bibinfo{journal}{Phys.Rev.Lett.} \textbf{\bibinfo{volume}{87}},
  \bibinfo{pages}{071301} (\bibinfo{year}{2001}), \eprint{nucl-ex/0106015}.

\bibitem[{\citenamefont{Ahmad et~al.}(2002)}]{Ahmad:2002jz}
\bibinfo{author}{\bibfnamefont{Q.}~\bibnamefont{Ahmad}} \bibnamefont{et~al.}
  (\bibinfo{collaboration}{SNO Collaboration}),
  \bibinfo{journal}{Phys.Rev.Lett.} \textbf{\bibinfo{volume}{89}},
  \bibinfo{pages}{011301} (\bibinfo{year}{2002}), \eprint{nucl-ex/0204008}.

\bibitem[{\citenamefont{An et~al.}(2012)}]{An:2012eh}
\bibinfo{author}{\bibfnamefont{F.}~\bibnamefont{An}} \bibnamefont{et~al.}
  (\bibinfo{collaboration}{DAYA-BAY Collaboration}),
  \bibinfo{journal}{Phys.Rev.Lett.} \textbf{\bibinfo{volume}{108}},
  \bibinfo{pages}{171803} (\bibinfo{year}{2012}), \eprint{1203.1669}.

\bibitem[{\citenamefont{Eguchi et~al.}(2003)}]{Eguchi:2002dm}
\bibinfo{author}{\bibfnamefont{K.}~\bibnamefont{Eguchi}} \bibnamefont{et~al.}
  (\bibinfo{collaboration}{KamLAND Collaboration}),
  \bibinfo{journal}{Phys.Rev.Lett.} \textbf{\bibinfo{volume}{90}},
  \bibinfo{pages}{021802} (\bibinfo{year}{2003}), \eprint{hep-ex/0212021}.

\bibitem[{\citenamefont{Ahn et~al.}(2006)}]{Ahn:2006zza}
\bibinfo{author}{\bibfnamefont{M.}~\bibnamefont{Ahn}} \bibnamefont{et~al.}
  (\bibinfo{collaboration}{K2K Collaboration}), \bibinfo{journal}{Phys.Rev.}
  \textbf{\bibinfo{volume}{D74}}, \bibinfo{pages}{072003}
  (\bibinfo{year}{2006}), \eprint{hep-ex/0606032}.

\bibitem[{\citenamefont{Abe et~al.}(2011)}]{Abe:2011sj}
\bibinfo{author}{\bibfnamefont{K.}~\bibnamefont{Abe}} \bibnamefont{et~al.}
  (\bibinfo{collaboration}{T2K Collaboration}),
  \bibinfo{journal}{Phys.Rev.Lett.} \textbf{\bibinfo{volume}{107}},
  \bibinfo{pages}{041801} (\bibinfo{year}{2011}), \eprint{1106.2822}.

\bibitem[{\citenamefont{Nakamura et~al.}(2010)\citenamefont{Nakamura, Hagiwara,
  Hikasa, Murayama, Tanabashi, Watari, Amsler, Antonelli, Asner, Baer
  et~al.}}]{1694801}
\bibinfo{author}{\bibfnamefont{K.}~\bibnamefont{Nakamura}},
  \bibinfo{author}{\bibfnamefont{K.}~\bibnamefont{Hagiwara}},
  \bibinfo{author}{\bibfnamefont{K.}~\bibnamefont{Hikasa}},
  \bibinfo{author}{\bibfnamefont{H.}~\bibnamefont{Murayama}},
  \bibinfo{author}{\bibfnamefont{M.}~\bibnamefont{Tanabashi}},
  \bibinfo{author}{\bibfnamefont{T.}~\bibnamefont{Watari}},
  \bibinfo{author}{\bibfnamefont{C.}~\bibnamefont{Amsler}},
  \bibinfo{author}{\bibfnamefont{M.}~\bibnamefont{Antonelli}},
  \bibinfo{author}{\bibfnamefont{D.~M.} \bibnamefont{Asner}},
  \bibinfo{author}{\bibfnamefont{H.}~\bibnamefont{Baer}}, \bibnamefont{et~al.},
  \textbf{\bibinfo{volume}{37}}, \bibinfo{pages}{1} (\bibinfo{year}{2010}),
  ISSN \bibinfo{issn}{0954-3899},
  \urlprefix\url{http://dx.doi.org/10.1088/0954-3899/37/7A/075021}.

\bibitem[{\citenamefont{{Abazajian} et~al.}(2013)\citenamefont{{Abazajian},
  {Arnold}, {Austermann}, {Benson}, {Bischoff}, {Bock}, {Bond}, {Borrill},
  {Calabrese}, {Carlstrom} et~al.}}]{CMBS4neu:2013}
\bibinfo{author}{\bibfnamefont{K.~N.} \bibnamefont{{Abazajian}}},
  \bibinfo{author}{\bibfnamefont{K.}~\bibnamefont{{Arnold}}},
  \bibinfo{author}{\bibfnamefont{J.}~\bibnamefont{{Austermann}}},
  \bibinfo{author}{\bibfnamefont{B.~A.} \bibnamefont{{Benson}}},
  \bibinfo{author}{\bibfnamefont{C.}~\bibnamefont{{Bischoff}}},
  \bibinfo{author}{\bibfnamefont{J.}~\bibnamefont{{Bock}}},
  \bibinfo{author}{\bibfnamefont{J.~R.} \bibnamefont{{Bond}}},
  \bibinfo{author}{\bibfnamefont{J.}~\bibnamefont{{Borrill}}},
  \bibinfo{author}{\bibfnamefont{E.}~\bibnamefont{{Calabrese}}},
  \bibinfo{author}{\bibfnamefont{J.~E.} \bibnamefont{{Carlstrom}}},
  \bibnamefont{et~al.}, \bibinfo{journal}{ArXiv e-prints}
  (\bibinfo{year}{2013}), \eprint{1309.5383}.

\bibitem[{\citenamefont{Ade et~al.}(2013)}]{Ade:2013zuv}
\bibinfo{author}{\bibfnamefont{P.}~\bibnamefont{Ade}} \bibnamefont{et~al.}
  (\bibinfo{collaboration}{Planck Collaboration}) (\bibinfo{year}{2013}),
  \eprint{1303.5076}.

\bibitem[{\citenamefont{Palanque-Delabrouille
  et~al.}(2014)\citenamefont{Palanque-Delabrouille, Yèche, Lesgourgues, Rossi,
  Borde et~al.}}]{Palanque-Delabrouille:2014jca}
\bibinfo{author}{\bibfnamefont{N.}~\bibnamefont{Palanque-Delabrouille}},
  \bibinfo{author}{\bibfnamefont{C.}~\bibnamefont{Yèche}},
  \bibinfo{author}{\bibfnamefont{J.}~\bibnamefont{Lesgourgues}},
  \bibinfo{author}{\bibfnamefont{G.}~\bibnamefont{Rossi}},
  \bibinfo{author}{\bibfnamefont{A.}~\bibnamefont{Borde}}, \bibnamefont{et~al.}
  (\bibinfo{year}{2014}), \eprint{1410.7244}.

\bibitem[{\citenamefont{Wyman et~al.}(2014)\citenamefont{Wyman, Rudd,
  Vanderveld, and Hu}}]{Wyman:2013lza}
\bibinfo{author}{\bibfnamefont{M.}~\bibnamefont{Wyman}},
  \bibinfo{author}{\bibfnamefont{D.~H.} \bibnamefont{Rudd}},
  \bibinfo{author}{\bibfnamefont{R.~A.} \bibnamefont{Vanderveld}},
  \bibnamefont{and} \bibinfo{author}{\bibfnamefont{W.}~\bibnamefont{Hu}},
  \bibinfo{journal}{Phys.Rev.Lett.} \textbf{\bibinfo{volume}{112}},
  \bibinfo{pages}{051302} (\bibinfo{year}{2014}), \eprint{1307.7715}.

\bibitem[{\citenamefont{Dvorkin et~al.}(2014)\citenamefont{Dvorkin, Wyman,
  Rudd, and Hu}}]{Dvorkin:2014lea}
\bibinfo{author}{\bibfnamefont{C.}~\bibnamefont{Dvorkin}},
  \bibinfo{author}{\bibfnamefont{M.}~\bibnamefont{Wyman}},
  \bibinfo{author}{\bibfnamefont{D.~H.} \bibnamefont{Rudd}}, \bibnamefont{and}
  \bibinfo{author}{\bibfnamefont{W.}~\bibnamefont{Hu}},
  \bibinfo{journal}{Phys.Rev.} \textbf{\bibinfo{volume}{D90}},
  \bibinfo{pages}{083503} (\bibinfo{year}{2014}), \eprint{1403.8049}.

\bibitem[{\citenamefont{Sunyaev and Zeldovich}(1980)}]{Sunyaev:1980nv}
\bibinfo{author}{\bibfnamefont{R.}~\bibnamefont{Sunyaev}} \bibnamefont{and}
  \bibinfo{author}{\bibfnamefont{Y.}~\bibnamefont{Zeldovich}},
  \bibinfo{journal}{Mon.Not.Roy.Astron.Soc.} \textbf{\bibinfo{volume}{190}},
  \bibinfo{pages}{413} (\bibinfo{year}{1980}).

\bibitem[{\citenamefont{Hand et~al.}(2012)\citenamefont{Hand, Addison, Aubourg,
  Battaglia, Battistelli et~al.}}]{Hand:2012ui}
\bibinfo{author}{\bibfnamefont{N.}~\bibnamefont{Hand}},
  \bibinfo{author}{\bibfnamefont{G.~E.} \bibnamefont{Addison}},
  \bibinfo{author}{\bibfnamefont{E.}~\bibnamefont{Aubourg}},
  \bibinfo{author}{\bibfnamefont{N.}~\bibnamefont{Battaglia}},
  \bibinfo{author}{\bibfnamefont{E.~S.} \bibnamefont{Battistelli}},
  \bibnamefont{et~al.}, \bibinfo{journal}{Phys.Rev.Lett.}
  \textbf{\bibinfo{volume}{109}}, \bibinfo{pages}{041101}
  (\bibinfo{year}{2012}), \eprint{1203.4219}.

\bibitem[{\citenamefont{Mueller et~al.}(2014)\citenamefont{Mueller,
  de~Bernardis, Bean, and Niemack}}]{Mueller:2014nsa}
\bibinfo{author}{\bibfnamefont{E.-M.} \bibnamefont{Mueller}},
  \bibinfo{author}{\bibfnamefont{F.}~\bibnamefont{de~Bernardis}},
  \bibinfo{author}{\bibfnamefont{R.}~\bibnamefont{Bean}}, \bibnamefont{and}
  \bibinfo{author}{\bibfnamefont{M.}~\bibnamefont{Niemack}}
  (\bibinfo{year}{2014}), \eprint{1408.6248}.

\bibitem[{\citenamefont{Lesgourgues and Pastor}(2006)}]{Lesgourgues:2006nd}
\bibinfo{author}{\bibfnamefont{J.}~\bibnamefont{Lesgourgues}} \bibnamefont{and}
  \bibinfo{author}{\bibfnamefont{S.}~\bibnamefont{Pastor}},
  \bibinfo{journal}{Phys.Rept.} \textbf{\bibinfo{volume}{429}},
  \bibinfo{pages}{307} (\bibinfo{year}{2006}), \eprint{astro-ph/0603494}.

\bibitem[{\citenamefont{Lesgourgues and Pastor}(2014)}]{Lesgourgues:2014zoa}
\bibinfo{author}{\bibfnamefont{J.}~\bibnamefont{Lesgourgues}} \bibnamefont{and}
  \bibinfo{author}{\bibfnamefont{S.}~\bibnamefont{Pastor}},
  \bibinfo{journal}{New J.Phys.} \textbf{\bibinfo{volume}{16}},
  \bibinfo{pages}{065002} (\bibinfo{year}{2014}), \eprint{1404.1740}.

\bibitem[{\citenamefont{{Abazajian} et~al.}(2011)\citenamefont{{Abazajian},
  {Calabrese}, {Cooray}, {De Bernardis}, {Dodelson}, {Friedland}, {Fuller},
  {Hannestad}, {Keating}, {Linder} et~al.}}]{2011APh....35..177A}
\bibinfo{author}{\bibfnamefont{K.~N.} \bibnamefont{{Abazajian}}},
  \bibinfo{author}{\bibfnamefont{E.}~\bibnamefont{{Calabrese}}},
  \bibinfo{author}{\bibfnamefont{A.}~\bibnamefont{{Cooray}}},
  \bibinfo{author}{\bibfnamefont{F.}~\bibnamefont{{De Bernardis}}},
  \bibinfo{author}{\bibfnamefont{S.}~\bibnamefont{{Dodelson}}},
  \bibinfo{author}{\bibfnamefont{A.}~\bibnamefont{{Friedland}}},
  \bibinfo{author}{\bibfnamefont{G.~M.} \bibnamefont{{Fuller}}},
  \bibinfo{author}{\bibfnamefont{S.}~\bibnamefont{{Hannestad}}},
  \bibinfo{author}{\bibfnamefont{B.~G.} \bibnamefont{{Keating}}},
  \bibinfo{author}{\bibfnamefont{E.~V.} \bibnamefont{{Linder}}},
  \bibnamefont{et~al.}, \bibinfo{journal}{Astroparticle Physics}
  \textbf{\bibinfo{volume}{35}}, \bibinfo{pages}{177} (\bibinfo{year}{2011}),
  \eprint{1103.5083}.

\bibitem[{\citenamefont{Lewis et~al.}(2000)\citenamefont{Lewis, Challinor, and
  Lasenby}}]{Lewis:1999bs}
\bibinfo{author}{\bibfnamefont{A.}~\bibnamefont{Lewis}},
  \bibinfo{author}{\bibfnamefont{A.}~\bibnamefont{Challinor}},
  \bibnamefont{and} \bibinfo{author}{\bibfnamefont{A.}~\bibnamefont{Lasenby}},
  \bibinfo{journal}{Astrophys.J.} \textbf{\bibinfo{volume}{538}},
  \bibinfo{pages}{473} (\bibinfo{year}{2000}), \eprint{astro-ph/9911177}.

\bibitem[{\citenamefont{Brandbyge et~al.}(2010)\citenamefont{Brandbyge,
  Hannestad, Haugboelle, and Wong}}]{Brandbyge:2010ge}
\bibinfo{author}{\bibfnamefont{J.}~\bibnamefont{Brandbyge}},
  \bibinfo{author}{\bibfnamefont{S.}~\bibnamefont{Hannestad}},
  \bibinfo{author}{\bibfnamefont{T.}~\bibnamefont{Haugboelle}},
  \bibnamefont{and} \bibinfo{author}{\bibfnamefont{Y.~Y.} \bibnamefont{Wong}},
  \bibinfo{journal}{JCAP} \textbf{\bibinfo{volume}{1009}}, \bibinfo{pages}{014}
  (\bibinfo{year}{2010}), \eprint{1004.4105}.

\bibitem[{\citenamefont{Castorina et~al.}(2014)\citenamefont{Castorina,
  Sefusatti, Sheth, Villaescusa-Navarro, and Viel}}]{Castorina:2013wga}
\bibinfo{author}{\bibfnamefont{E.}~\bibnamefont{Castorina}},
  \bibinfo{author}{\bibfnamefont{E.}~\bibnamefont{Sefusatti}},
  \bibinfo{author}{\bibfnamefont{R.~K.} \bibnamefont{Sheth}},
  \bibinfo{author}{\bibfnamefont{F.}~\bibnamefont{Villaescusa-Navarro}},
  \bibnamefont{and} \bibinfo{author}{\bibfnamefont{M.}~\bibnamefont{Viel}},
  \bibinfo{journal}{JCAP} \textbf{\bibinfo{volume}{1402}}, \bibinfo{pages}{049}
  (\bibinfo{year}{2014}), \eprint{1311.1212}.

\bibitem[{\citenamefont{{Bhattacharya}
  et~al.}(2011)\citenamefont{{Bhattacharya}, {Heitmann}, {White}, {Luki{\'c}},
  {Wagner}, and {Habib}}}]{2011ApJ...732..122B}
\bibinfo{author}{\bibfnamefont{S.}~\bibnamefont{{Bhattacharya}}},
  \bibinfo{author}{\bibfnamefont{K.}~\bibnamefont{{Heitmann}}},
  \bibinfo{author}{\bibfnamefont{M.}~\bibnamefont{{White}}},
  \bibinfo{author}{\bibfnamefont{Z.}~\bibnamefont{{Luki{\'c}}}},
  \bibinfo{author}{\bibfnamefont{C.}~\bibnamefont{{Wagner}}}, \bibnamefont{and}
  \bibinfo{author}{\bibfnamefont{S.}~\bibnamefont{{Habib}}},
  \bibinfo{journal}{\apj} \textbf{\bibinfo{volume}{732}}, \bibinfo{eid}{122}
  (\bibinfo{year}{2011}), \eprint{1005.2239}.

\bibitem[{\citenamefont{Kiakotou et~al.}(2008)\citenamefont{Kiakotou, Elgaroy,
  and Lahav}}]{Kiakotou:2007pz}
\bibinfo{author}{\bibfnamefont{A.}~\bibnamefont{Kiakotou}},
  \bibinfo{author}{\bibfnamefont{O.}~\bibnamefont{Elgaroy}}, \bibnamefont{and}
  \bibinfo{author}{\bibfnamefont{O.}~\bibnamefont{Lahav}},
  \bibinfo{journal}{Phys.Rev.} \textbf{\bibinfo{volume}{D77}},
  \bibinfo{pages}{063005} (\bibinfo{year}{2008}), \eprint{0709.0253}.

\bibitem[{\citenamefont{Linder}(2005)}]{Linder:2005in}
\bibinfo{author}{\bibfnamefont{E.~V.} \bibnamefont{Linder}},
  \bibinfo{journal}{Phys.Rev.} \textbf{\bibinfo{volume}{D72}},
  \bibinfo{pages}{043529} (\bibinfo{year}{2005}), \eprint{astro-ph/0507263}.

\bibitem[{\citenamefont{Sheth et~al.}(2001)\citenamefont{Sheth, Diaferio, Hui,
  and Scoccimarro}}]{Sheth:2000ff}
\bibinfo{author}{\bibfnamefont{R.~K.} \bibnamefont{Sheth}},
  \bibinfo{author}{\bibfnamefont{A.}~\bibnamefont{Diaferio}},
  \bibinfo{author}{\bibfnamefont{L.}~\bibnamefont{Hui}}, \bibnamefont{and}
  \bibinfo{author}{\bibfnamefont{R.}~\bibnamefont{Scoccimarro}},
  \bibinfo{journal}{Mon.Not.Roy.Astron.Soc.} \textbf{\bibinfo{volume}{326}},
  \bibinfo{pages}{463} (\bibinfo{year}{2001}), \eprint{astro-ph/0010137}.

\bibitem[{\citenamefont{Bhattacharya and Kosowsky}(2008)}]{Bhattacharya:2007sk}
\bibinfo{author}{\bibfnamefont{S.}~\bibnamefont{Bhattacharya}}
  \bibnamefont{and} \bibinfo{author}{\bibfnamefont{A.}~\bibnamefont{Kosowsky}},
  \bibinfo{journal}{Phys.Rev.} \textbf{\bibinfo{volume}{D77}},
  \bibinfo{pages}{083004} (\bibinfo{year}{2008}), \eprint{0712.0034}.

\bibitem[{\citenamefont{Battaglia et~al.}(2010)\citenamefont{Battaglia, Bond,
  Pfrommer, Sievers, and Sijacki}}]{Battaglia:2010tm}
\bibinfo{author}{\bibfnamefont{N.}~\bibnamefont{Battaglia}},
  \bibinfo{author}{\bibfnamefont{J.}~\bibnamefont{Bond}},
  \bibinfo{author}{\bibfnamefont{C.}~\bibnamefont{Pfrommer}},
  \bibinfo{author}{\bibfnamefont{J.}~\bibnamefont{Sievers}}, \bibnamefont{and}
  \bibinfo{author}{\bibfnamefont{D.}~\bibnamefont{Sijacki}},
  \bibinfo{journal}{Astrophys.J.} \textbf{\bibinfo{volume}{725}},
  \bibinfo{pages}{91} (\bibinfo{year}{2010}), \eprint{1003.4256}.

\bibitem[{\citenamefont{Sazonov and Sunyaev}(1999)}]{Sazonov:1999zp}
\bibinfo{author}{\bibfnamefont{S.}~\bibnamefont{Sazonov}} \bibnamefont{and}
  \bibinfo{author}{\bibfnamefont{R.}~\bibnamefont{Sunyaev}},
  \bibinfo{journal}{Mon.Not.Roy.Astron.Soc.} \textbf{\bibinfo{volume}{310}},
  \bibinfo{pages}{765} (\bibinfo{year}{1999}), \eprint{astro-ph/9903287}.

\bibitem[{\citenamefont{Laureijs et~al.}(2011)}]{Laureijs:2011gra}
\bibinfo{author}{\bibfnamefont{R.}~\bibnamefont{Laureijs}} \bibnamefont{et~al.}
  (\bibinfo{collaboration}{EUCLID Collaboration}) (\bibinfo{year}{2011}),
  \eprint{1110.3193}.

\bibitem[{\citenamefont{Rykoff et~al.}(2014)}]{Rykoff:2013ovv}
\bibinfo{author}{\bibfnamefont{E.}~\bibnamefont{Rykoff}} \bibnamefont{et~al.}
  (\bibinfo{collaboration}{SDSS}), \bibinfo{journal}{Astrophys.J.}
  \textbf{\bibinfo{volume}{785}}, \bibinfo{pages}{104} (\bibinfo{year}{2014}),
  \eprint{1303.3562}.

\bibitem[{\citenamefont{Niemack et~al.}(2010)\citenamefont{Niemack, Ade,
  Aguirre, Barrientos, Beall et~al.}}]{Niemack:2010wz}
\bibinfo{author}{\bibfnamefont{M.}~\bibnamefont{Niemack}},
  \bibinfo{author}{\bibfnamefont{P.}~\bibnamefont{Ade}},
  \bibinfo{author}{\bibfnamefont{J.}~\bibnamefont{Aguirre}},
  \bibinfo{author}{\bibfnamefont{F.}~\bibnamefont{Barrientos}},
  \bibinfo{author}{\bibfnamefont{J.}~\bibnamefont{Beall}},
  \bibnamefont{et~al.}, \bibinfo{journal}{Proc.SPIE Int.Soc.Opt.Eng.}
  \textbf{\bibinfo{volume}{7741}}, \bibinfo{pages}{77411S}
  (\bibinfo{year}{2010}), \eprint{1006.5049}.

\bibitem[{\citenamefont{Dawson et~al.}(2012)}]{Dawson:2012va}
\bibinfo{author}{\bibfnamefont{K.~S.} \bibnamefont{Dawson}}
  \bibnamefont{et~al.} (\bibinfo{collaboration}{BOSS Collaboration})
  (\bibinfo{year}{2012}), \eprint{1208.0022}.

\bibitem[{\citenamefont{Calabrese et~al.}(2014)\citenamefont{Calabrese,
  Hložek, Battaglia, Bond, de~Bernardis et~al.}}]{Calabrese:2014gwa}
\bibinfo{author}{\bibfnamefont{E.}~\bibnamefont{Calabrese}},
  \bibinfo{author}{\bibfnamefont{R.}~\bibnamefont{Hložek}},
  \bibinfo{author}{\bibfnamefont{N.}~\bibnamefont{Battaglia}},
  \bibinfo{author}{\bibfnamefont{J.~R.} \bibnamefont{Bond}},
  \bibinfo{author}{\bibfnamefont{F.}~\bibnamefont{de~Bernardis}},
  \bibnamefont{et~al.}, \bibinfo{journal}{JCAP}
  \textbf{\bibinfo{volume}{1408}}, \bibinfo{pages}{010} (\bibinfo{year}{2014}),
  \eprint{1406.4794}.

\bibitem[{\citenamefont{Levi et~al.}(2013)}]{Levi:2013gra}
\bibinfo{author}{\bibfnamefont{M.}~\bibnamefont{Levi}} \bibnamefont{et~al.}
  (\bibinfo{collaboration}{DESI collaboration}) (\bibinfo{year}{2013}),
  \eprint{1308.0847}.

\bibitem[{\citenamefont{Lesgourgues et~al.}(2006)\citenamefont{Lesgourgues,
  Perotto, Pastor, and Piat}}]{Lesgourgues:2005yv}
\bibinfo{author}{\bibfnamefont{J.}~\bibnamefont{Lesgourgues}},
  \bibinfo{author}{\bibfnamefont{L.}~\bibnamefont{Perotto}},
  \bibinfo{author}{\bibfnamefont{S.}~\bibnamefont{Pastor}}, \bibnamefont{and}
  \bibinfo{author}{\bibfnamefont{M.}~\bibnamefont{Piat}},
  \bibinfo{journal}{Phys.Rev.} \textbf{\bibinfo{volume}{D73}},
  \bibinfo{pages}{045021} (\bibinfo{year}{2006}), \eprint{astro-ph/0511735}.

\bibitem[{\citenamefont{{Okamoto} and {Hu}}(2003)}]{2003PhRvD..67h3002O}
\bibinfo{author}{\bibfnamefont{T.}~\bibnamefont{{Okamoto}}} \bibnamefont{and}
  \bibinfo{author}{\bibfnamefont{W.}~\bibnamefont{{Hu}}},
  \bibinfo{journal}{\prd} \textbf{\bibinfo{volume}{67}}, \bibinfo{eid}{083002}
  (\bibinfo{year}{2003}), \eprint{astro-ph/0301031}.

\bibitem[{\citenamefont{Keisler and Schmidt}(2013)}]{Keisler:2012eg}
\bibinfo{author}{\bibfnamefont{R.}~\bibnamefont{Keisler}} \bibnamefont{and}
  \bibinfo{author}{\bibfnamefont{F.}~\bibnamefont{Schmidt}},
  \bibinfo{journal}{Astrophys.J.} \textbf{\bibinfo{volume}{765}},
  \bibinfo{pages}{L32} (\bibinfo{year}{2013}), \eprint{1211.0668}.

\bibitem[{\citenamefont{Font-Ribera et~al.}(2014)\citenamefont{Font-Ribera,
  McDonald, Mostek, Reid, Seo et~al.}}]{Font-Ribera:2013rwa}
\bibinfo{author}{\bibfnamefont{A.}~\bibnamefont{Font-Ribera}},
  \bibinfo{author}{\bibfnamefont{P.}~\bibnamefont{McDonald}},
  \bibinfo{author}{\bibfnamefont{N.}~\bibnamefont{Mostek}},
  \bibinfo{author}{\bibfnamefont{B.~A.} \bibnamefont{Reid}},
  \bibinfo{author}{\bibfnamefont{H.-J.} \bibnamefont{Seo}},
  \bibnamefont{et~al.}, \bibinfo{journal}{JCAP}
  \textbf{\bibinfo{volume}{1405}}, \bibinfo{pages}{023} (\bibinfo{year}{2014}),
  \eprint{1308.4164}.

\end{thebibliography}

\end{document}